\documentclass[manuscript]{geophysics}

\usepackage{multirow}
\usepackage{amsmath}

\usepackage{algorithm,algorithmic}

\usepackage[colorlinks,
linkcolor=blue,
urlcolor = blue,
citecolor=blue
]{hyperref}

\usepackage{hyperref}

\begin{document}

\title{An anti-noise seismic inversion method based on diffusion model}

\renewcommand{\thefootnote}{\fnsymbol{footnote}} 


\address{
\footnotemark[1]BP UTG, \\
200 Westlake Park Blvd, \\
Houston, TX, 77079 \\
\footnotemark[2]Bureau of Economic Geology, \\
John A. and Katherine G. Jackson School of Geosciences \\
The University of Texas at Austin \\
University Station, Box X \\
Austin, TX 78713-8924}
\author{Joe Dellinger\footnotemark[1] and Sergey Fomel\footnotemark[2]}

\address{
	\footnotemark[1]Key Laboratory of Earth Exploration and Information Technology of Ministry of Education, Chengdu University of Technology, China. E-mail: yingtianliu06@outlook.cn\\
	\footnotemark[2]State Key Lab of Oil and Gas Reservoir Geology and Exploitation; School of Geophysics, Chengdu University of Technology, China. E-mail: liyong07@cdut.edu.cn(corresponding author); 2390328914@qq.com; zhangquanliao@outlook.com; 454810391@qq.com;\\
}

\author{Liu Yingtian\footnotemark[1], Li Yong\footnotemark[2], Li Huating\footnotemark[2], Liao Zhangquan\footnotemark[2], Feng Wen\footnotemark[2].}

\lefthead{Liu et al.} 
\righthead{The DSIM-USSL} 

\maketitle

\begin{abstract}

Seismic impedance inversion is one of the most important part of geophysical exploration. However, due to random noise, the traditional semi-supervised learning (SSL) methods lack generalization and stability. To solve this problem, some authors have proposed SSL methods with anti-noise function to improve noise robustness and inversion accuracy. However, such methods are often not ideal when faced with strong noise. In addition, Low-frequency impedance models can mitigate this problem, but creating accurate low-frequency models is difficult and error-prone when well-log data is sparse and subsurface structures are complex. To address those issues, we propose a novel deep learning inversion method called DSIM-USSL (Unsupervised and Semi-supervised joint Learning for Seismic Inversion based on Diffusion Model). Specifically, we are the first to introduce a diffusion model with strong noise tendency and construct a diffusion seismic inversion model (DSIM). In the reverse diffusion of DSIM, we design the encoder-decoder which combines CNN for capturing local features and GRU for global sequence modeling; and we choose U-net to learn the distribution of random noise, enhancing the generalization and stability of proposed method. Furthermore, to further improve generalization of the proposed method, a two-step training approach (USSL) is utilized. First, an unsupervised trained encoder-decoder is used as the initial network model in place of the traditional low-frequency wave impedance model that is difficult to accurately acquire. Then, the SSL is employed to further optimize the encoder-decoder model. Experimental results on the Marmousi2 model and field data demonstrate that the DSIM-USSL method achieves higher accuracy in the presence of seismic data with random noise, and maintains high stability even under strong noise conditions.

\end{abstract}

\section{Introduction}

Seismic impedance inversion is an important technique for evaluating and characterizing reservoirs, which has great significance in seismic exploration. In this technique, a large amount of seismic data, including amplitude and travel time of seismic signals, is widely used for inverting various physical parameters such as subsurface lithology and hydrocarbon-bearing reservoirs. However, due to the unknown seismic wavelets, limited bandwidth of observed data, and various interferences from noise sources, the inversion problem in seismic imaging is often ill-posed, nonlinear, and non-unique. To overcome these limitations, regularization can be performed by imposing certain constraints on the solution space. The solution objective function is then solved using a stochastic or deterministic approach \citep{duijndam1988bayesian,buland2003bayesian,gholami2015nonlinear}. These methods include least-squares inversion\citep{wright1986new}, Bayesian inversion\citep{de2017bayesian}, and global optimization algorithms\citep{bijani2017physical}. However, due to the complexity of geological structures, environment and saturated fluids, it is challenging to accurately establish the forward operator and initial inversion model. Consequently, effective solutions to corresponding geophysical inversion problems are difficult to achieve.

In recent years, with the development of artificial intelligence technology, deep learning has become a mainstream technique for data analysis. The concept was first proposed by Hinton et al. in the field of neural network research\citep{hinton2006fast}. Traditional methods that can only learn single-layer features without hierarchical structure, such as clustering algorithms\citep{kanungo2002efficient,xu2015comprehensive} support vector machines\citep{cherkassky2004practical,wang2005comparison}, and maximum entropy methods\citep{jaynes1982rationale,el2015arabic}. Deep learning, on the other hand, learns hierarchical feature representations of data through layer-by-layer feature transformations\citep{ojala2002multiresolution,lowe2004distinctive}. Meanwhile, many models have emerged in the field of deep learning, such as convolutional neural networks\citep{gu2018recent} (CNN), deep belief networks\citep{hinton2006reducing} (DBN), auto-encoder\citep{wang2016auto} (AE), generative adversarial network\citep{goodfellow2020generative} (GAN), capsule network\citep{mazzia2021efficient} (CapsNet), and attention mechanism model Transformer\citep{jaderberg2015spatial}. These models have achieved great success in several domains such as computer vision\citep{chai2021deep}, image processing\citep{wang2020deep}, speech recognition\citep{li2022recent}, and natural language processing\citep{min2023recent} due to their powerful ability to abstract and extract high-level semantic features. Compared to traditional non-deep learning methods, deep learning has been popularized and successfully applied in different fields such as computer science, information engineering and geophysics. With the introduction of deep learning, many problems have been solved in geophysics, including fault identification\citep{lin2022automatic} salt body detection\citep{guo2020deep}, seismic phase classification\citep{wu2023seismic}, reservoir prediction\citep{song2022reservoir}, and seismic denoising\citep{li2023multiple}. Meanwhile, various nonlinear inversion problems in geophysics have been developed, such as transverse wave velocity modeling\citep{wang2022s,yang2022reconstruction}, three-parameter inversion\citep{li2022pertinent,liu2023multi}, and seismic full waveform inversion\citep{he2021reparameterized,zhang2021deep}. In recent years, researchers have explored a lot of supervised learning algorithms for wave impedance inversion, and Das et al. used CNN to obtain the geophysical problem of subsurface elasticity modeling from recorded positive-incidence seismic data\citep{das2019convolutional}. Mustafa et al. used TCN sequence modeling to predict the wave impedance from the seismic data, which overcame the problems of vanishing gradient of the RNN and overfitting of the CNN\citep{mustafa2019estimation}. However, in practice, the generalization ability of the network is limited due to the limited amount of well-labeled data\citep{dou2023contrasinver}.

To improve the generalization ability of the network, some researchers have proposed semi-supervised learning (SSL) frameworks for impedance inversion\citep{alfarraj2019semisupervised,wu2021semi,ning2024transformer}, which improves the generalization ability of the wave impedance deep learning inversion network from a training perspective. However, such SSL methods increase the dependence on seismic data while improving network generalization, and when seismic data is noisy, the stability of SSL inversion results is poor. To improve the stability of SSL inversion, low-frequency model constraints can be added\citep{zhang2021robust}, but in practical applications, the common method for establishing an initial model is well-log data interpolation\citep{zhang2023structurally}. When well-log data is sparse and underground structures are complex, creating an accurate initial model is often error-prone\citep{zou2023domain}. To handle random noise in SSL inversion, anti-noise network structures are typically adopted, such as the Noisy Softplus (NSP) activation function\citep{liu2017noisy} and the Rand Softplus (RSP) adaptive noise activation function \citep{chen2019improving}, which have achieved good results in practical applications. However, when the noise level is too high, the stability of the inversion result is still poor.

The diffusion model (DM) is a model based on Bayesian likelihood probability, inspired by non-equilibrium thermodynamics\citep{rezende2015variational,ho2020denoising,rombach2022high,hoogeboom2021autoregressive}. Compared to the mainstream generative adversarial networks (GAN)\citep{goodfellow2020generative}, variational autoencoders (VAE)\citep{lopez2018information}, flow-based models\citep{ho2019flow++}, and energy-based models (EBM)\citep{du2019implicit}, the diffusion model is less prone to mode collapse and training failure. By leveraging the nonlinear characteristics of the diffusion equation, the diffusion model can progressively remove random noise and interference from seismic signals, thereby extracting the true seismic signal. This effectively reduces the impact of random noise on seismic data processing and interpretation, improving the quality and reliability of seismic data. Moreover, the diffusion model does not require prior information about the subsurface medium; it only needs the input of raw seismic data.

\section{Theory and methodology}

\subsection{Diffusion seismic inversion model}

The diffusion model is mainly divided into forward process and reverse process, as shown in Figure~\ref{fig:figure1}. In the forward process, the original data is destroyed by gradually adding Gaussian noise. $\beta_{t}$ is a pre-defined diffusion rate, where $t$ gradually increases from time $1$ to $T$. To simulate the speed increase in the actual diffusion process, it is expressed as:
\begin{equation}\label{H1}
\beta_{1} < \beta_{2} < \cdots \beta_{t}.
\end{equation}
Therefore, the image $x_{t}$ of seismic data after adding noise can be expressed as:
\begin{equation}\label{H2}
x_{t}=\sqrt{\beta_{t}}\cdot\epsilon_{t}+\sqrt{1-\beta_{t}}\cdot x_{t-1},  
\end{equation}
where $x_{0}$ represents the image input from the true data set. $x_{t}$ represents the image after adding $t$ noise to $x_{0}$. $\epsilon_{t}$ represents the random noise added at the $t$ time and satisfies $\epsilon_{t} \sim \mathcal N(0,1)$. To simplify the derivation, we set $\alpha_{t}=1-\beta_{t}$ From equation ~\ref{H2} , it can be deduced:
\begin{equation}\label{H3}
x_{t}=\sqrt{1-\alpha_{t}}\cdot\epsilon_{t}+\sqrt{\alpha_{t}}\cdot x_{t-1},  
\end{equation}
\begin{equation}\label{H4}
x_{t-1}=\sqrt{1-\alpha_{t-1}}\cdot\epsilon_{t-1}+\sqrt{\alpha_{t-1}}\cdot x_{t-2}. 
\end{equation}
According to equation~\ref{H3} and equation~\ref{H4}, $x_{t}$ can also be expressed as:
\begin{equation}\label{H5}
x_{t}=\sqrt{1-\alpha_{t}}\cdot\epsilon_{t}+\sqrt{\alpha_{t}}\cdot \sqrt{1-\alpha_{t-1}}\cdot\epsilon_{t-1}+\sqrt{\alpha_{t}}\cdot\sqrt{\alpha_{t-1}}\cdot x_{t-2}.
\end{equation}
Because $\epsilon_{t}$ is a normal distribution, we can introduce the reparameterization technique. Equation~\ref{H5} can be expressed as:
\begin{equation}\label{H6}
x_{t}=\sqrt{1-\alpha_{t}\alpha_{t-1}}\cdot\epsilon+\sqrt{\alpha_{t}\alpha_{t-1}}\cdot x_{t-2}.
\end{equation}
Furthermore, equation~\ref{H6} is repeatedly reversed, and can be derived as follows:
\begin{equation}\label{H7}
x_{t}=\sqrt{1-\overline{\alpha_{t}}}\cdot\epsilon + \sqrt{\overline{\alpha_{t}}}\cdot x_{0},  
\end{equation}
where $\overline{\alpha_{t}}=\prod_{i=1}^{T}\alpha_{i}$. Therefore, the image $x_{t}$ is only related to $x_{t-1}$ at the previous time. The forward process is represented by $q\left(x_{t}\middle| x_{t-1}\right)$, which can be regarded as a Markov process:
\begin{equation}\label{H8}
q\left(x_{1:T}\middle| x_{0}\right)=\prod_{t=1}^{T}{q\left(x_{t}\middle|\ x_{t-1}\right)},
\end{equation}
\begin{equation}\label{H9}
q\left(x_{t}\middle| x_{t-1}\right)= \mathcal N\left(x_{t},\sqrt{1-\beta_{t}}x_{t-1},\beta_{t}I\right), 
\end{equation}
where $T$ represents the number of times noise is added. When $T$ is large enough, $x_T$ can be regarded as a normal distribution. According to equation~\ref{H8} and equation~\ref{H9}, the relationship between $x_t$ and $x_0$ can be expressed as:
\begin{equation}\label{H10}
q\left(x_t\middle| x_0\right)= \mathcal N\left(x_t;\sqrt{\overline{\alpha_t}}x_0,\left(1-\overline{\alpha_t}\right)I\right).
\end{equation}
In the reverse process, two tasks are designed to be learned. To recover the original input data from the noisy data, the first task is used to learn the process of denoise. The reverse process is represented by $q(x_{t-1}|x_t)$, which can be expressed as:
\begin{equation}\label{H11}
q\left(x_{t-1}\middle| x_t,x_0\right)=\mathcal N\left(x_{t-1}\middle|\widetilde{\mu_t}\left(x_t,x_0\right),\widetilde{\beta_t}I\right),
\end{equation}
where $\widetilde{\mu_t}$ and $\widetilde{\beta_t}$ are the seismic image mean and noise distribution at time $t$, respectively. Assuming that a network is used instead of the reverse process, it can be expressed as:
\begin{equation}\label{H12}
p_\theta\left(x_{t-1}\middle| x_t\right)=\mathcal N\left(x_{t-1}\middle|\mu_\theta\left(x_t,t\right),\sum_{\theta}\left(x_t,t\right)\right),
\end{equation}
where $p_\theta\left(x_{t-1}\middle| x_t\right)$ represents the reverse process of network prediction. $\mu_\theta\left(x_t,t\right)$ and $\sum_{\theta}\left(x_t,t\right)$ represent the mean and equation of the network prediction, respectively. Because the variance of the Gaussian distribution in the forward process has been obtained. For the reverse process, we only need to fit $\mu_\theta$ using the neural network. Furthermore, according to equation~\ref{H11} and equation~\ref{H12}, the mathematical relation between $\mu_\theta$ and $\epsilon_{\theta\left(x_t,t\right)}$ can be expressed as:
\begin{equation}\label{H13}
\mu_\theta\left(x_t,t\right)=\frac{1}{\sqrt{\alpha_t}}\left(x_t-\frac{1-\alpha_t}{\sqrt{1-\overline{\alpha_t}}}\epsilon_{\theta\left(x_t,t\right)}\right).
\end{equation}
Because $t$ and $x_t$ are already obtained, we only need to fit $\epsilon_{\theta(x_t,t)}$. In view of the high efficiency of image feature extraction, the U-net network is used to train the denoising diffusion model.

The second task was used to learn the process of seismic inversion, unsupervised learning to train the seismic inversion model. The network model is expressed as:
\begin{equation}\label{H14}
Forward(\mathcal{F}_\omega\left(x_{t-1}\right))\approx x_{t-1},
\end{equation}
where $\mathcal{F}_\omega$ represents the seismic inversion network. It can be expressed as $\mathcal{F}_\omega=[\omega_\mathcal{F}^1,\omega_\mathcal{F}^2,\ldots,\omega_\mathcal{F}^{L-1},\omega_\mathcal{F}^L]$, and $\omega_\mathcal{F}^i$ is the weight vector of layer $i$ in the network structure. $Forward$ represents the forward modeling module, which can be expressed as:
\begin{equation}\label{H15}
Forward(n)=w\left(n\right)\ast R\left(n\right),
\end{equation}
where $n$ is the number of stratigraphic sequences, $w\left(n\right)$ is the wavelet function, and $\ast$ is the symbol of the convolution. $R(n)$ is the reflection coefficient, which can be expressed as:
\begin{equation}\label{H16}
R(n)=\frac{I_n-I_{n-1}}{I_n+I_{n-1}},
\end{equation}
where $I_n$ represents the wave impedance of the $n$ layer.

\subsection{Seismic Inversion module}
The seismic inversion module in the DSIM model will be used many times. Therefore, to improve computational efficiency and ensure the long-term dependence of the model, we use CNN to extract local features and Gated Recurrent Unit (GRU) for global sequence modeling. The GRU network is a variant recurrent neural network (RNN) \citep{cho2014learning}. It solves the problems of gradient disappearance and gradient explosion when using RNN to deal with long-term dependence. Furthermore, the GRU structure is simpler and easier to implement than the LSTM with a three-gate structure\citep{yang2020lstm,shewalkar2018comparison}. The structure of GRU network is shown in Figure~\ref{fig:figure2}, which mainly consists of reset gate $r_t$ and update gate $z_t$. It can be expressed in mathematical equations as:
\begin{equation}\label{H17}
\left\{\begin{matrix}
 r_t=\sigma (W_r\cdot [h_{t-1},x_t])\\
 z_t=\sigma (W_z\cdot [h_{t-1},x_t])\\
 \widetilde{h_t} = tanh(W\cdot [r_1 \cdot  h_{t-1},x_t])\\
 h_t = (1-z_t) \cdot h_{t-1} + z_t \cdot \widetilde{h_t}
\end{matrix}\right.,
\end{equation}
where $W_r$ and $W_z$ represent the weight matrix of the update and reset gates, respectively. $x_t$ is the input information. $h_{t-1}$, ${\widetilde{h}}_t$, and $h_t$ represent the upper hidden state, the candidate hidden state, and current hidden state, respectively. With the input $x_t$ and $h_{t-1}$, the sigmoid activation function is used to calculate the values of the reset and update gates. Then, combined with the reset gate calculated in the previous step, the tanh activation function is used to calculate the candidate hidden state ${\widetilde{h}}_t$. Next, the update gate is used as a weight coefficient, and $h_t$ is obtained by adding ${\widetilde{h}}_t$ and $h_{t-1}$ by weight.

In addition, seismic inversion is not only a sequence regression problem, but also a typical encoding to decoding problem. Therefore, an encoder-decoder for inversion is constructed, as shown in Figure~\ref{fig:figure3}. There are two paths in the process of data coding: one for sequence modeling through the GRU layer, and the other for learning global features through the CNN layer. The output of these two paths is combined by channel and fed to the decoder. The decoder consists of an upsampling layer and a regression layer. Upsampling layer is used to match the seismic data with the wave impedance data sampling rate, while the regression layer is used to output the wave impedance data. It is worth noting that through the encoder-decoder, global and local information of seismic data is fully extracted, and finally high-precision wave impedance is obtained.

\subsection{Unsupervised and Semi-supervised joint Learning}
To improve the generalization of inversion, we integrate the extraction of the initial model into the training. A step-by-step training method combining unsupervised and semi-supervised learning (USSL) is designed, as shown in Figure~\ref{fig:figure4}. First, we pre-train the seismic inversion module with unsupervised learning to get the initial network model. It is used to replace low-frequency models that are difficult to obtain accurately in practice. Then, it is applied to the target seismic area by transfer learning. Finally, to optimize the parameters of the network model, the seismic data and well-log data of the target area are used as semi-supervised learning.

The pre-training process of the source seismic area is described in Algorithm~\ref{algorithm1}. The denoising diffusion network model $\epsilon_\theta$ and the initial network model $\mathcal{F}_{\omega_1}$ are obtained by Algorithm~\ref{algorithm1}. To prevent excessive noise of input data, network training will lose its meaning\citep{song2022learning,fawzi2017robustness}. The training threshold m is introduced into the seismic inversion module. The training process of the target seismic area is described in Algorithm~\ref{algorithm2}. The seismic inversion module $\mathcal{F}_{\omega_2}$ is obtained by semi-supervised learning training. The loss function weighting of semi-supervised learning in Algorithm~\ref{algorithm2} can be expressed as:
\begin{equation}\label{H18}
Loss=k_u\cdot {Loss}_u+k_s\cdot {Loss}_s,
\end{equation}
where ${Loss}_u$ is the Loss function of unsupervised learning, ${Loss}_s$ is the loss function of supervised learning. $k_u$ and $k_s$ are the weight coefficients of ${Loss}_u$ and ${Loss}_s$, respectively.

\begin{algorithm}[!ht]
	\caption{Pre-training the model}
    \label{algorithm1}
    \begin{algorithmic}[1] 
        \STATE $\textbf{Input:}$ $\mathit{T}$, Number of noises added; $m$, Training threshold\\
        $x_{0}\sim q(x_{0})$ and $t\sim$ Uniform($\left \{ 1,...,T \right \}$)\\
        \STATE $\epsilon \sim \mathcal N(0,1)$\\
        \FOR {$t=1,\dots,T$}
        \STATE $\nabla _ { \theta } | | \epsilon - \epsilon _ { \theta } ( \sqrt { \overline { \alpha } _ { t } } x _ { 0 } + \sqrt { 1 - \overline { \alpha } _ { t } } \epsilon , t ) | | ^ { 2 } $
                \IF {$t \le m$}
                \STATE $Forward(\mathcal{F}_{\omega_{1}}(x_{t-1}))  \approx x_{t-1}$\\
                \ENDIF
        \ENDFOR
    \RETURN $\epsilon _ { \theta }$, $\mathcal{F}_{\omega_{1}}$
    \end{algorithmic}
\end{algorithm}

\begin{algorithm}[!ht]
	\caption{Training the model} 
    \label{algorithm2}
    \begin{algorithmic}[1] 
        \STATE $\textbf{Input:}$ $\mathit{T}$: Number of noises added; $\epsilon _ { \theta }$; $\mathcal{F}_{\omega_{1}}$\\
        $x_{T}\sim \mathcal N(0,1)$\\
        \FOR {$t=T,\dots,1$}
             \STATE $z\sim \mathcal N(0,1)$ if $t>1$, else $z=0$\\
             \STATE $x _ { t - 1 } = \frac { 1 } { \sqrt { \alpha _ { t } } } ( x _ { t } - \frac { 1 - \alpha _ { t } } { \sqrt { 1 - \overline { \alpha } _ { t } } } f _ { \theta } ( x _ { t } , t ) ) + \sqrt { 1 - \alpha _ { t } } z $\\
              \IF {$t=1$}
                \STATE $Forward(\mathcal{F}_{\omega_{2}}(x_{t-1}))  \approx x_{t-1}$\\
                \WHILE{well-log data $m_{i}$ exists}
                    \STATE $\mathcal{F}_{\omega_{2}} (x_{t-1})\approx m_{i}$\\                    
                \ENDWHILE
              \ENDIF        
        \ENDFOR
    \RETURN $\mathcal{F}_{\omega_{2}}$
    \end{algorithmic}
\end{algorithm}

\section{Numerical experiments} 

\subsection{2D Model data set generation} 

To evaluate the generalization and random noise robustness of the proposed method, we conducted tests on the Marmousi2 model. The marmousi2 model is a two-dimensional synthetic model developed by the French Institute of Petroleum Geosciences. It is widely utilized for seismic data processing and interpretation. For additional details, please refer to Reference~\citep{Martin2006Marmousi2}. First, we use the marmousi2 model to synthesize the reflection coefficient. Then, the seismic data is obtained by convolution of the reflection coefficient with a main frequency of 40Hz. Furthermore, to improve the speed of network convergence, the seismic data is normalized by the equation:
\begin{equation}\label{H19}
x^\prime=\frac{x-\bar{x}}{\sigma\left(x\right)},
\end{equation}
where $x$ represents the original seismic data, $\bar{x}$ and $\sigma(x)$ are the mean and standard deviation of $x$, respectively, and $x^\prime$ represents the normalized seismic data. Additionally, the seismic data was downsampled by a factor of 4 to simulate the resolution disparity between it and the well-log data. 

The synthetic seismic data profile was divided into 18 smaller profiles arranged in 6 rows and 3 columns. Four profiles, highlighted with gray fill, were designated for training, and the remaining profiles were allocated for pre-training, as shown in Figure~\ref{fig:figure5}(a). Figure~\ref{fig:figure5}(b)-(e) shows the four seismic profiles that were used in training and testing, each spanning a horizontal range of 1600 meters and a depth of 320 meters. Specifically, Figure~\ref{fig:figure5}(b)-(c) are seismic profiles of areas with reservoir development, which is used to assess the influence of the method on the reservoir. Figure~\ref{fig:figure5}(d)-(e) shows seismic profiles with significant amplitude changes and developed faults, which are used to assess the applicability of the method in complex formations.

\subsection{Hyperparameter Settings and performance metrics} 

During pre-training, we set the number of the DSIM noises to 2000, the number of iterations to 1000, the learning rate and the batch size are set to 2e-5 and 16. In the seismic inversion module, both the coded GRU layers and the convolutional block channels are set to 8. The training threshold is set to 1, with the number of iterations at 2. The initial learning rate is set to 0.005, weight decay to 0.0001, and the dropout value to 0.2, using the Adam optimizer for network optimization. During training, the batch size for codec-decoder iterations was set to 50, with a total of 500 iterations. It takes only 2 minutes to run on a computer equipped with the Inter i5-12490f and a single Nvidia GeForce GTX 3060 GPU.

To quantitatively evaluate the performance of the proposed workflow, we used Pearson correlation coefficient (PCC), Coefficient of determination (R²) and Structural similarity, SSIM). Specifically, the PCC is used to measure the linear relationship between two variables. The R² is used to evaluate the goodness of fit of the regression model. The SSIM is used to compare the structural similarity of two images. Higher values of these indicators signify stronger correlations and thus better model performance. They are defined as:
\begin{equation}\label{H20}
PCC=\frac{\sum_{i=1}^{N}\left(y_i-\mu_y\right)\left({\hat{y}}_i-\mu_{\hat{y}}\right)}{\sqrt{\sum_{i=1}^{N}\left(y_i-\mu_y\right)^2}\sqrt{\sum_{i=1}^{n}\left({\hat{y}}_i-\mu_{\hat{y}}\right)^2}},
\end{equation}
\begin{equation}\label{H21}
R^2=1-\frac{\sum_{i=1}^{N}{(y_i-{\hat{y}}_i)^2}}{\sum_{i=1}^{N}{(y_i-\mu_y)^2}},
\end{equation}
\begin{equation}\label{H22}
SSIM=\frac{\left(2\mu_y\mu_{\hat{y}}+C_1\right)\left(2\sigma_{y\hat{y}}+C_2\right)}{\left(\mu_y^2+\mu_{\hat{y}}^2+C_1\right)\left(\sigma_y^2+\sigma_{\hat{y}}^2+C_2\right)},
\end{equation}
where $y$ and $\hat{y}$ represent the true value and predicted value, respectively. $\mu_y$ and $\mu_{\hat{y}}$ represent the mean value of $y$ and $\hat{y}$, respectively. $\sigma_{y\hat{y}}$ represents the covariance between $y$ and $\hat{y}$. $C_1$ and $C_2$ represent constants, and they are usually set to a smaller positive value to avoid zeros in the denominator.

\subsection{Synthetic data experiment} 

To investigate the impact of USSL training methods on DSIM, we compared it with training methods of UL, SL and SSL. The initial model of the network was pre-trained by 14 seismic profiles. Subsequently, the initial model was transferred to the designated four test seismic profiles. We used only 12 impedance traces for training to simulate well-log data in a practical setting, which is far less than 1$\%$ of the total number of traces. In addition, the seismic data of the entire seismic trace is used as an input for unsupervised learning. The inversion results derived from various training methods are presented in Figure~\ref{fig:figure6}. Although the UL method can characterize the boundarie of the four seismic profiles, the inversion results significantly deviate from the true model, as shown in Figure~\ref{fig:figure6}(a)(f)(k)(p). The accuracy of the SL method significantly surpasses that of the UL method, as shown in Figure~\ref{fig:figure6}(b)(g)(l)(q). However, the SL method exhibits outliers around the reservoir region and complex formations, such as noticeable artifacts at the orange arrow in Figure~\ref{fig:figure6}(g) and Figure~\ref{fig:figure6}(i). Compared with the SL method, the inversion results of the SSL and USSL methods more closely resemble the true model, as shown in columns three and four of Figure~\ref{fig:figure6}. Specifically, in certain areas, the USSL method can more accurately capture the wave impedance details, notably those delineated by the rectangles in Figure~\ref{fig:figure6}(r) and Figure~\ref{fig:figure6}(s).

Figure~\ref{fig:figure7} shows the loss function curves of different methods in the four seismic profiles. All methods converge smoothly during training, as shown in Figure~\ref{fig:figure7}(a)-(d). Although the UL method achieves the fastest convergence rate in training, it exhibits the worst in the validation sets. This is mainly due to the UL method focuses on minimizing the difference between the input and the reconstructed seismic data. Unlike the UL method, the SL, SSL, and USSL models aim to minimize the difference between the predicted and true wave impedance. Consequently, the SL, SSL, and USSL demonstrate superior fitting to labeled data during training. It is worth noting that the UL method converges at the beginning of training, as shown in Figure~\ref{fig:figure7}(e)-(h) where the red arrows are located. Therefore, we need to select a small number of iterations to pre-train the initial network model in the USSL method. The top-right rectangular box in Figure~\ref{fig:figure7}(e)-(h) shows the magnification of the validation loss function curve for the last 100 iterations. It can be seen that USSL achieves higher accuracy, followed by the SSL method.

Moreover, the results of unlabeled single trace inversion in the four seismic profiles are shown in Figure~\ref{fig:figure8}. Four unlabeled single traces are extracted from each seismic profile. Notably, the inversion results of the UL method are omitted from the figure~\ref{fig:figure8} due to significant oscillations. The SL method accurately fits the true results in general. However, the red arrow indicates a deviation from the true model. It's even a standard deviation away from the true model. Compared to the SL method, both the SSL and USSL methods demonstrate improved performance. Specifically, the USSL method surpasses the SSL method in the region highlighted by the yellow arrow. To quantitatively assess the inversion effect of each method, PCC, R2, and SSIM were calculated. Each method has 12 index values and the results are summarized as shown in Table~\ref{tbl:table1}. The 11 optimal results were obtained by the USSL method. Therefore, the USSL method has positive effects on the quality of inversion results in DSIM-USSL.

To examine the robustness of the proposed method over random noise, we compare it with three SSL methods: unconditional SSL, SSL employing the noisy softplus activation function (NSP-SSL), and SSL utilizing the noise adaptive activation function rand softplus (RSP-SSL). Random noise was added to the original seismic data at various signal-to-noise ratio (SNR). Five sets of experiments were designed with SNR of -5, -2, 0, 2, and 5. Among them, three sets were designated as low-SNR conditions to evaluate the algorithm's performance under strong random noise. The inversion results of the four seismic profiles are shown in Figure~\ref{fig:figure9}. The first row shows the raw seismic data with noise, with left-to-right SNR of -5, -2, 0, 2, and 5. The second to fifth rows shows the inversion results of the unconditional SSL method, NSP-SSL method, RSP-SSL method, and DSIM-USSL method, respectively. It is worth noting that as the noise level increases, the inversion effect of the unconditional SSL method exhibits high instability and a significant decrease. Both NSP-SSL and RSP-SSL methods can effectively restrain the impact of noise in high SNR environments. However, their performance significantly decreases under strong noise. Compared to other SSL methods, the DSIM-USSL method provides a clearer delineation of reservoir and formation boundaries, even in environments with high levels of noise. Figure~\ref{fig:figure10} shows the absolute difference between the inversion result and the true value. It is observed that as noise increases, the reservoir location DSIM-USSL method is almost unaffected by noise, as shown by the yellow arrow in Figure~\ref{fig:figure10}(a). Furthermore, the DSIM-USSL method is also significantly better than other methods in the case of strong noise, as shown by the yellow arrow in figure~\ref{fig:figure10}(b). Even at the interface of complex formations, the absolute difference of DSIM-USSL is least affected by noise, the area surrounded by the rectangular box in Figure~\ref{fig:figure10}(c) and Figure~\ref{fig:figure10}(d).

The validation loss curves of each method are shown in Figure~\ref{fig:figure11}, Figure~\ref{fig:figure12}, and Figure~\ref{fig:figure13}. where the vertical axis represents the loss of the mean square error. The disturbed regions of the loss of the DSIM-USSL method are all below the unconditional SSL, as shown in Figure ~\ref{fig:figure11}. It is worth noting that the DSIM-USSL method is better at SNR=-5 than SSL at SNR=5. Therefore, the DSIM-USSL method is completely superior to the unconditional SSL method. The right rectangular area in Figure~\ref{fig:figure11} shows the magnification of the last 100 iterations of the DSIM-USSL method. Compared to the shadow range of SSL method and DSIM-USSL method, the DSIM-USSL method is at least three times less affected by noise than the unconditional SSL method. Figure~\ref{fig:figure12} and Figure~\ref{fig:figure13} show the loss of the DSIM-USSL method compared to the SSL method of noise resistance. In simple geological environments, the NSP-SSL and RSP-SSL methods demonstrate reduced sensitivity to noise, as shown in Figures~\ref{fig:figure12}(a)-(b). However, under complex geological conditions, the performance of NSP-SSL and RSP-SSL methods show great volatility. In comparison, the DSIM-USSL method shows markedly better stability, which is particularly evident in Figure~\ref{fig:figure12}(d) and Figure~\ref{fig:figure13}(d).

Furthermore, we conduct a quantitative comparison of the inversion results from each method, and the dot plots of PCC, R² and SSIM are shown in Figure~\ref{fig:figure14}(a)-(c). The PCC and SSIM scores for the  DSIM-USSL methods generally surpass those of the unconditional SSL, NSP-SSL, and RSP-SSL methods across all noise scenarios. It is slightly lower than the NSP-SSL method and the RSP-SSL method, only in the case of the fourth seismic profile with a SNR of 5. The R² coefficient of the DSIM-USSL method is absolutely superior in the first three seismic profiles. Moreover, the DSIM-USSL method achieves the highest R² coefficients in three out of five SNR categories for the fourth seismic profile. Therefore, the DSIM-USSL method demonstrates enhanced robustness to random noise and achieves higher accuracy compared to conventional SSL methods.

\section{Field Data Example} 
To assess the applicability of the proposed method, we utilized field data from the H Oilfield in China. The field is rich in tight sandstone reservoirs, but the quality of seismic data is poor due to complex surface conditions. Considerable effort was devoted to denoising seismic data from the H Oilfield. Figure~\ref{fig:figure15}(a)-(d) shows four distinct seismic data profiles obtained after the denoising process. The time range and the offset range of seismic data are 0.0–1.2s and 0–720 m, and the time sampling interval is 1ms. The initial network model was obtained by pre-training on four seismic profiles. Subsequently, the model was applied to the seismic profiles without noise processing for further training, as shown in Figure~\ref{fig:figure16}(b). In addition, the seismic profile includes wells W1, W2, and W3 with CDP of 60m, 174m, and 240m, respectively, as shown in Figure~\ref{fig:figure16}(b). wells W1 and W3 are used for training, while well W2 are used for testing. We obtained the Ricker theoretical wavelet with a main frequency of 28Hz by scanning and smoothing the seismic track, as shown in Figure~\ref{fig:figure16}(a).

Figure~\ref{fig:figure17} shows the wave impedance results of different inversion methods. The results of the unconditional SSL, NSP-SSL, and RSP-SSL are shown in Figure~\ref{fig:figure17}(a), Figure~\ref{fig:figure17}(b) and, Figure~\ref{fig:figure17}(c), respectively. Figure 17(c) shows the results of the proposed method. It can be noted that there is a sandstone gas reservoir at the time of 0.4s, as shown by the yellow arrow in Figure~\ref{fig:figure17}. Due to the existence of noise, the reservoir obtained by unconditional SSL method exhibits considerable indistinctness. The NSP-SSL and RSP-SSL methods provide clearer results, but still lack accuracy in reservoir definition. Compared with these SSL methods, the DSIM-USSL method has a better performance in revealing the lateral changes of gas layers. In addition, the DSIM-USSL method also has better performance in stratigraphic division. As shown in the area of the oval wire frame in Figure~\ref{fig:figure17}, sandstone with a strip shape is predicted by the DSIM-USSL method. Moreover, the inversion results of each method in well W2 are shown in Figure~\ref{fig:figure18}. It can be noted that the prediction results of each method are within the standard deviation of the true value. However, the arrows areas in Figure~\ref{fig:figure18} show that the unconditional SSL method, NSP-SSL method, and RSP-SSL method have large errors. The PCC, R² and SSIM evaluation indexes of well W2 are shown in Table~\ref{tbl:table2}. Among them, the DSIM-USSL method has the best PCC, R² and SSIM indexes.

\section{Conclusion}

In this article, a novel seismic inversion method is proposed, called DSIM-USSL. It is designed to overcome the influence of random noise on seismic inversion results. The DSIM is derived using a diffusion mechanism, which can actively remove random noise in the inversion. Furthermore, a two-step training style of unsupervised learning and semi-supervised learning combination (USSL) was designed. The acquisition of the initial network model is integrated into the training. Both experiments on the synthetic data show that DSIM-USSI outperforms the traditional unconditional SSL method and SSL methods with anti-noise function in seismic inversion. The 
inversion results of field data match well with the filtered well-log data, which proves the feasibility of this method. In addition, the proposed method could also be used in other SSL inversion fields, such as elastic impedance inversion, full waveform inversion, and three-parameter inversion; these all require high quality seismic data and accurate low-frequency models.

\section{Data and materials availability} 
Data related to this study can be obtained by contacting the corresponding author.

\bibliographystyle{seg}  
\bibliography{DSIM-USSL}

\begin{thebibliography}{}
\itemsep0pt

\bibitem[Alfarraj and AlRegib, 2019]{alfarraj2019semisupervised}
Alfarraj, M., and G. AlRegib,  2019, Semisupervised sequence modeling for elastic impedance inversion: Interpretation, {\bfseries 7}, SE237--SE249.

\bibitem[Bijani et~al., 2017]{bijani2017physical}
Bijani, R., P.~G. Leli{\`e}vre, C.~F. Ponte-Neto, and C.~G. Farquharson,  2017, Physical-property-, lithology-and surface-geometry-based joint inversion using pareto multi-objective global optimization: Geophysical Journal International, {\bfseries 209}, 730--748.

\bibitem[Buland and Omre, 2003]{buland2003bayesian}
Buland, A., and H. Omre,  2003, Bayesian linearized avo inversion: Geophysics, {\bfseries 68}, 185--198.

\bibitem[Chai et~al., 2021]{chai2021deep}
Chai, J., H. Zeng, A. Li, and E.~W. Ngai,  2021, Deep learning in computer vision: A critical review of emerging techniques and application scenarios: Machine Learning with Applications, {\bfseries 6}, 100134.

\bibitem[Chen et~al., 2019]{chen2019improving}
Chen, Y., Y. Mai, J. Xiao, and L. Zhang,  2019, Improving the antinoise ability of dnns via a bio-inspired noise adaptive activation function rand softplus: Neural computation, {\bfseries 31}, 1215--1233.

\bibitem[Cherkassky and Ma, 2004]{cherkassky2004practical}
Cherkassky, V., and Y. Ma,  2004, Practical selection of svm parameters and noise estimation for svm regression: Neural networks, {\bfseries 17}, 113--126.

\bibitem[Cho et~al., 2014]{cho2014learning}
Cho, K., B. Van~Merri{\"e}nboer, C. Gulcehre, D. Bahdanau, F. Bougares, H. Schwenk, and Y. Bengio,  2014, Learning phrase representations using rnn encoder-decoder for statistical machine translation: arXiv preprint arXiv:1406.1078.

\bibitem[Das et~al., 2019]{das2019convolutional}
Das, V., A. Pollack, U. Wollner, and T. Mukerji,  2019, Convolutional neural network for seismic impedance inversion: Geophysics, {\bfseries 84}, R869--R880.

\bibitem[de~Figueiredo et~al., 2017]{de2017bayesian}
de~Figueiredo, L.~P., D. Grana, M. Santos, W. Figueiredo, M. Roisenberg, and G.~S. Neto,  2017, Bayesian seismic inversion based on rock-physics prior modeling for the joint estimation of acoustic impedance, porosity and lithofacies: Journal of Computational Physics, {\bfseries 336}, 128--142.

\bibitem[Dou et~al., 2023]{dou2023contrasinver}
Dou, Y., T. Li, K. Li, H. Duan, and Z. Xu,  2023, Contrasinver: Voxel-wise contrastive semi-supervised learning for seismic inversion: arXiv preprint arXiv:2302.06441.

\bibitem[Du and Mordatch, 2019]{du2019implicit}
Du, Y., and I. Mordatch,  2019, Implicit generation and modeling with energy based models: Advances in Neural Information Processing Systems, {\bfseries 32}.

\bibitem[Duijndam, 1988]{duijndam1988bayesian}
Duijndam, A.,  1988, Bayesian estimation in seismic inversion. part i: Principles 1: Geophysical Prospecting, {\bfseries 36}, 878--898.

\bibitem[El-Halees, 2015]{el2015arabic}
El-Halees, A.~M.,  2015, Arabic text classification using maximum entropy: IUG Journal of Natural Studies, {\bfseries 15}.

\bibitem[Fawzi et~al., 2017]{fawzi2017robustness}
Fawzi, A., S.-M. Moosavi-Dezfooli, and P. Frossard,  2017, The robustness of deep networks: A geometrical perspective: IEEE Signal Processing Magazine, {\bfseries 34}, 50--62.

\bibitem[Gholami, 2015]{gholami2015nonlinear}
Gholami, A.,  2015, Nonlinear multichannel impedance inversion by total-variation regularization: Geophysics, {\bfseries 80}, R217--R224.

\bibitem[Goodfellow et~al., 2020]{goodfellow2020generative}
Goodfellow, I., J. Pouget-Abadie, M. Mirza, B. Xu, D. Warde-Farley, S. Ozair, A. Courville, and Y. Bengio,  2020, Generative adversarial networks: Communications of the ACM, {\bfseries 63}, 139--144.

\bibitem[Gu et~al., 2018]{gu2018recent}
Gu, J., Z. Wang, J. Kuen, L. Ma, A. Shahroudy, B. Shuai, T. Liu, X. Wang, G. Wang, J. Cai, et~al.,  2018, Recent advances in convolutional neural networks: Pattern recognition, {\bfseries 77}, 354--377.

\bibitem[Guo et~al., 2020]{guo2020deep}
Guo, J., L. Xu, J. Ding, B. He, S. Dai, and F. Liu,  2020, A deep supervised edge optimization algorithm for salt body segmentation: IEEE Geoscience and Remote Sensing Letters, {\bfseries 18}, 1746--1750.

\bibitem[He and Wang, 2021]{he2021reparameterized}
He, Q., and Y. Wang,  2021, Reparameterized full-waveform inversion using deep neural networks: Geophysics, {\bfseries 86}, V1--V13.

\bibitem[Hinton et~al., 2006]{hinton2006fast}
Hinton, G.~E., S. Osindero, and Y.-W. Teh,  2006, A fast learning algorithm for deep belief nets: Neural computation, {\bfseries 18}, 1527--1554.

\bibitem[Hinton and Salakhutdinov, 2006]{hinton2006reducing}
Hinton, G.~E., and R.~R. Salakhutdinov,  2006, Reducing the dimensionality of data with neural networks: science, {\bfseries 313}, 504--507.

\bibitem[Ho et~al., 2019]{ho2019flow++}
Ho, J., X. Chen, A. Srinivas, Y. Duan, and P. Abbeel,  2019, Flow++: Improving flow-based generative models with variational dequantization and architecture design: International Conference on Machine Learning, PMLR, 2722--2730.

\bibitem[Ho et~al., 2020]{ho2020denoising}
Ho, J., A. Jain, and P. Abbeel,  2020, Denoising diffusion probabilistic models: Advances in neural information processing systems, {\bfseries 33}, 6840--6851.

\bibitem[Hoogeboom et~al., 2021]{hoogeboom2021autoregressive}
Hoogeboom, E., A.~A. Gritsenko, J. Bastings, B. Poole, R.~v.~d. Berg, and T. Salimans,  2021, Autoregressive diffusion models: arXiv preprint arXiv:2110.02037.

\bibitem[Jaderberg et~al., 2015]{jaderberg2015spatial}
Jaderberg, M., K. Simonyan, A. Zisserman, et~al.,  2015, Spatial transformer networks: Advances in neural information processing systems, {\bfseries 28}.

\bibitem[Jaynes, 1982]{jaynes1982rationale}
Jaynes, E.~T.,  1982, On the rationale of maximum-entropy methods: Proceedings of the IEEE, {\bfseries 70}, 939--952.

\bibitem[Kanungo et~al., 2002]{kanungo2002efficient}
Kanungo, T., D.~M. Mount, N.~S. Netanyahu, C.~D. Piatko, R. Silverman, and A.~Y. Wu,  2002, An efficient k-means clustering algorithm: Analysis and implementation: IEEE transactions on pattern analysis and machine intelligence, {\bfseries 24}, 881--892.

\bibitem[Li et~al., 2022a]{li2022recent}
Li, J., et~al.,  2022a, Recent advances in end-to-end automatic speech recognition: APSIPA Transactions on Signal and Information Processing, {\bfseries 11}.

\bibitem[Li et~al., 2023]{li2023multiple}
Li, J., R. Qu, and C. Lu,  2023, Multiple attention mechanisms-based convolutional neural network for desert seismic denoising: Pure and Applied Geophysics,  1--21.

\bibitem[Li et~al., 2022b]{li2022pertinent}
Li, Z., X. Chen, J. Li, and J. Zhang,  2022b, Pertinent multigate mixture-of-experts-based prestack three-parameter seismic inversion: IEEE Transactions on Geoscience and Remote Sensing, {\bfseries 60}, 1--15.

\bibitem[Lin et~al., 2022]{lin2022automatic}
Lin, L., Z. Zhong, Z. Cai, A.~Y. Sun, and C. Li,  2022, Automatic geologic fault identification from seismic data using 2.5 d channel attention u-net: Geophysics, {\bfseries 87}, IM111--IM124.

\bibitem[Liu et~al., 2017]{liu2017noisy}
Liu, Q., Y. Chen, and S. Furber,  2017, Noisy softplus: an activation function that enables snns to be trained as anns: arXiv preprint arXiv:1706.03609.

\bibitem[Liu et~al., 2023]{liu2023multi}
Liu, X., B. Wu, and H. Yang,  2023, Multi-task full attention u-net for prestack seismic inversion: IEEE Geoscience and Remote Sensing Letters.

\bibitem[Lopez et~al., 2018]{lopez2018information}
Lopez, R., J. Regier, M.~I. Jordan, and N. Yosef,  2018, Information constraints on auto-encoding variational bayes: Advances in neural information processing systems, {\bfseries 31}.

\bibitem[Lowe, 2004]{lowe2004distinctive}
Lowe, D.~G.,  2004, Distinctive image features from scale-invariant keypoints: International journal of computer vision, {\bfseries 60}, 91--110.

\bibitem[Martin et~al., 2006]{Martin2006Marmousi2}
Martin, G.~S., R. Wiley, and K.~J. Marfurt,  2006, Marmousi2: An elastic upgrade for marmousi: The leading edge, {\bfseries 25}, 156--166.

\bibitem[Mazzia et~al., 2021]{mazzia2021efficient}
Mazzia, V., F. Salvetti, and M. Chiaberge,  2021, Efficient-capsnet: Capsule network with self-attention routing: Scientific reports, {\bfseries 11}, 14634.

\bibitem[Min et~al., 2023]{min2023recent}
Min, B., H. Ross, E. Sulem, A.~P.~B. Veyseh, T.~H. Nguyen, O. Sainz, E. Agirre, I. Heintz, and D. Roth,  2023, Recent advances in natural language processing via large pre-trained language models: A survey: ACM Computing Surveys, {\bfseries 56}, 1--40.

\bibitem[Mustafa et~al., 2019]{mustafa2019estimation}
Mustafa, A., M. Alfarraj, and G. AlRegib,  2019, Estimation of acoustic impedance from seismic data using temporal convolutional network, {\itshape in} SEG Technical Program Expanded Abstracts 2019: Society of Exploration Geophysicists,  2554--2558.

\bibitem[Ning et~al., 2024]{ning2024transformer}
Ning, C., B. Wu, and B. Wu,  2024, Transformer and convolutional hybrid neural network for seismic impedance inversion: IEEE Journal of Selected Topics in Applied Earth Observations and Remote Sensing.

\bibitem[Ojala et~al., 2002]{ojala2002multiresolution}
Ojala, T., M. Pietikainen, and T. Maenpaa,  2002, Multiresolution gray-scale and rotation invariant texture classification with local binary patterns: IEEE Transactions on pattern analysis and machine intelligence, {\bfseries 24}, 971--987.

\bibitem[Rezende and Mohamed, 2015]{rezende2015variational}
Rezende, D., and S. Mohamed,  2015, Variational inference with normalizing flows: International conference on machine learning, PMLR, 1530--1538.

\bibitem[Rombach et~al., 2022]{rombach2022high}
Rombach, R., A. Blattmann, D. Lorenz, P. Esser, and B. Ommer,  2022, High-resolution image synthesis with latent diffusion models: Proceedings of the IEEE/CVF conference on computer vision and pattern recognition, 10684--10695.

\bibitem[Shewalkar, 2018]{shewalkar2018comparison}
Shewalkar, A.~N.,  2018, Comparison of rnn, lstm and gru on speech recognition data.

\bibitem[Song et~al., 2022a]{song2022reservoir}
Song, C., W. Lu, Y. Wang, S. Jin, J. Tang, and L. Chen,  2022a, Reservoir prediction based on closed-loop cnn and virtual well-logging labels: IEEE Transactions on Geoscience and Remote Sensing, {\bfseries 60}, 1--12.

\bibitem[Song et~al., 2022b]{song2022learning}
Song, H., M. Kim, D. Park, Y. Shin, and J.-G. Lee,  2022b, Learning from noisy labels with deep neural networks: A survey: IEEE Transactions on Neural Networks and Learning Systems.

\bibitem[Wang and Hu, 2005]{wang2005comparison}
Wang, H., and D. Hu,  2005, Comparison of svm and ls-svm for regression: 2005 International conference on neural networks and brain, IEEE, 279--283.

\bibitem[Wang et~al., 2022]{wang2022s}
Wang, J., J. Cao, S. Zhao, and Q. Qi,  2022, S-wave velocity inversion and prediction using a deep hybrid neural network: Science China Earth Sciences,  1--18.

\bibitem[Wang et~al., 2016]{wang2016auto}
Wang, Y., H. Yao, and S. Zhao,  2016, Auto-encoder based dimensionality reduction: Neurocomputing, {\bfseries 184}, 232--242.

\bibitem[Wang et~al., 2020]{wang2020deep}
Wang, Z., J. Chen, and S.~C. Hoi,  2020, Deep learning for image super-resolution: A survey: IEEE transactions on pattern analysis and machine intelligence, {\bfseries 43}, 3365--3387.

\bibitem[Wright and Holt, 1986]{wright1986new}
Wright, S., and J. Holt,  1986, A new non-linear least squares algorithm for the seismic inversion problem: Geophysical Journal International, {\bfseries 87}, 1041--1056.

\bibitem[Wu et~al., 2021]{wu2021semi}
Wu, B., D. Meng, and H. Zhao,  2021, Semi-supervised learning for seismic impedance inversion using generative adversarial networks: Remote Sensing, {\bfseries 13}, 909.

\bibitem[Wu et~al., 2023]{wu2023seismic}
Wu, H., S. Li, and N. Liu,  2023, Seismic interpolation via multi-scale hu-net: Geoenergy Science and Engineering, {\bfseries 222}, 211458.

\bibitem[Xu and Tian, 2015]{xu2015comprehensive}
Xu, D., and Y. Tian,  2015, A comprehensive survey of clustering algorithms: Annals of Data Science, {\bfseries 2}, 165--193.

\bibitem[Yang et~al., 2022]{yang2022reconstruction}
Yang, J., C. Xu, and Y. Zhang,  2022, Reconstruction of the s-wave velocity via mixture density networks with a new rayleigh wave dispersion function: IEEE Transactions on Geoscience and Remote Sensing, {\bfseries 60}, 1--13.

\bibitem[Yang et~al., 2020]{yang2020lstm}
Yang, S., X. Yu, and Y. Zhou,  2020, Lstm and gru neural network performance comparison study: Taking yelp review dataset as an example: 2020 International workshop on electronic communication and artificial intelligence (IWECAI), IEEE, 98--101.

\bibitem[Zhang et~al., 2021]{zhang2021robust}
Zhang, J., J. Li, X. Chen, Y. Li, G. Huang, and Y. Chen,  2021, Robust deep learning seismic inversion with a priori initial model constraint: Geophysical Journal International, {\bfseries 225}, 2001--2019.

\bibitem[Zhang and Gao, 2021]{zhang2021deep}
Zhang, W., and J. Gao,  2021, Deep-learning full-waveform inversion using seismic migration images: IEEE Transactions on Geoscience and Remote Sensing, {\bfseries 60}, 1--18.

\bibitem[Zhang et~al., 2023]{zhang2023structurally}
Zhang, Y., H. Zhou, M. Zhang, Y. Wang, B. Feng, and M. Liang,  2023, Structurally constrained initial impedance modeling for poststack seismic inversion: IEEE Transactions on Geoscience and Remote Sensing, {\bfseries 61}, 1--10.

\bibitem[Zou et~al., 2023]{zou2023domain}
Zou, B., Y. Wang, T. Chen, J. Liang, G. Yu, and G. Hu,  2023, The domain adversarial and spatial fusion semi-supervised seismic impedance inversion: IEEE Transactions on Geoscience and Remote Sensing, {\bfseries 62}, 1--15.

\end{thebibliography}

\tabl{table1}{Performance of inversion results for different training methods.}
{
\centering
\begin{tabular}{|c|c|c|c|c|}
\hline
\textbf{Test Model}                & \textbf{Train scheme} & \textbf{PCC}    & \textbf{R²}     & \textbf{SSIM}   \\ \hline
\multirow{4}{*}{\textbf{Profile1}} & UL                    & 0.4020          & -1.0585         & 0.2327          \\ \cline{2-5} 
                                   & SL                    & 0.9121          & 0.8148          & 0.7862          \\ \cline{2-5} 
                                   & SSL                   & 0.9490          & 0.8911          & 0.8421          \\ \cline{2-5} 
                                   & USSL                  & \textbf{0.9577} & \textbf{0.9075} & \textbf{0.8714} \\ \hline
\multirow{4}{*}{\textbf{Profile2}} & UL                    & 0.7036          & 0.3263          & 0.3996          \\ \cline{2-5} 
                                   & SL                    & 0.9882          & 0.9751          & 0.8026          \\ \cline{2-5} 
                                   & SSL                   & 0.9924          & 0.9842          & 0.8410          \\ \cline{2-5} 
                                   & USSL                  & \textbf{0.9936} & \textbf{0.9866} & \textbf{0.8645} \\ \hline
\multirow{4}{*}{\textbf{Profile3}} & UL                    & 0.6541          & -0.3852         & 0.2367          \\ \cline{2-5} 
                                   & SL                    & 0.9394          & 0.7479          & 0.7066          \\ \cline{2-5} 
                                   & SSL                   & 0.9730          & 0.9295          & 0.7697          \\ \cline{2-5} 
                                   & USSL                  & \textbf{0.9790} & \textbf{0.9509} & \textbf{0.8151} \\ \hline
\multirow{4}{*}{\textbf{Profile4}} & UL                    & 0.6367          & -1.7567         & 0.2134          \\ \cline{2-5} 
                                   & SL                    & 0.8573          & 0.5858          & 0.5850          \\ \cline{2-5} 
                                   & SSL                   & \textbf{0.9262} & 0.7246          & 0.6084          \\ \cline{2-5} 
                                   & USSL                  & 0.9259          & \textbf{0.7432} & \textbf{0.6196} \\ \hline
\end{tabular}
}

\tabl{table2}{Performance of different inversion results at the location of the well W2 on the field data.}
{
\centering
\begin{center}
\begin{tabular}{|c|c|c|c|}
\hline
\textbf{Train scheme} & \textbf{PCC}    & \textbf{R²}     & \textbf{SSIM}   \\ \hline
Unconditional-SSL     & 0.9508          & 0.8853          & 0.6125          \\ \hline
NSP-SSL               & 0.9596          & 0.9154          & 0.6838          \\ \hline
RSP-SSL               & 0.9641          & 0.9176          & 0.6557          \\ \hline
DSIM-USSL             & \textbf{0.9832} & \textbf{0.9404} & \textbf{0.8607} \\ \hline
\end{tabular}
\end{center}
}

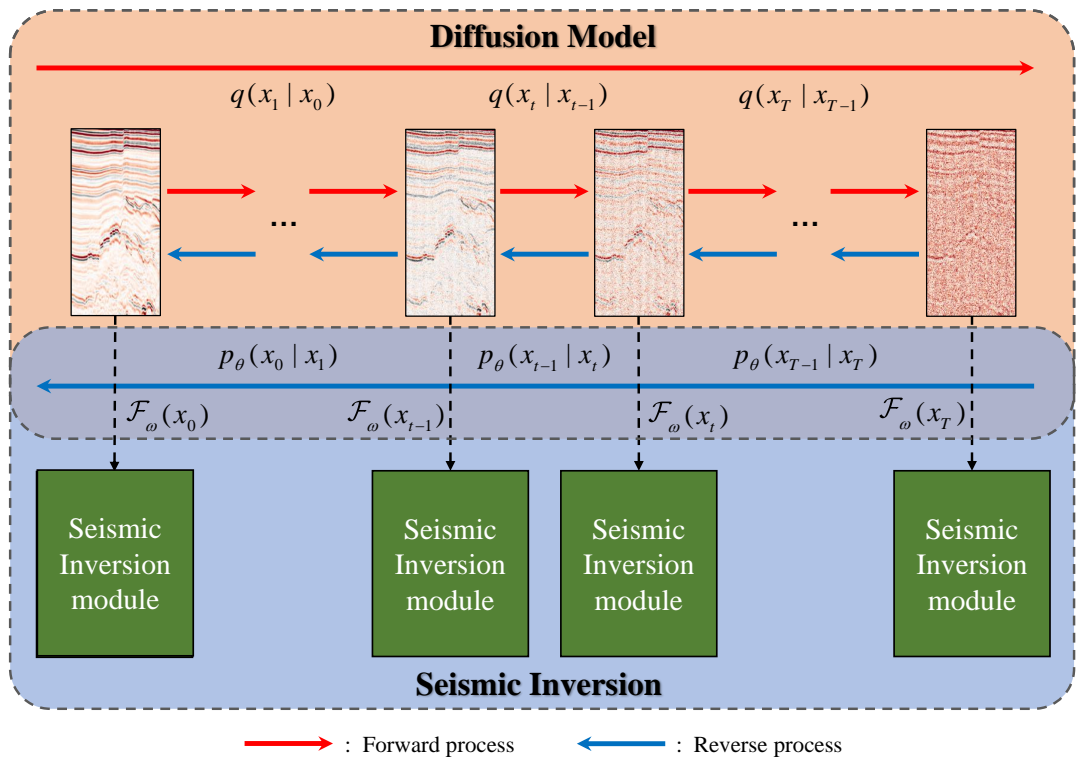
\plot{figure1}{width=1\textwidth}{The structure of the DSIM model.}

\newpage
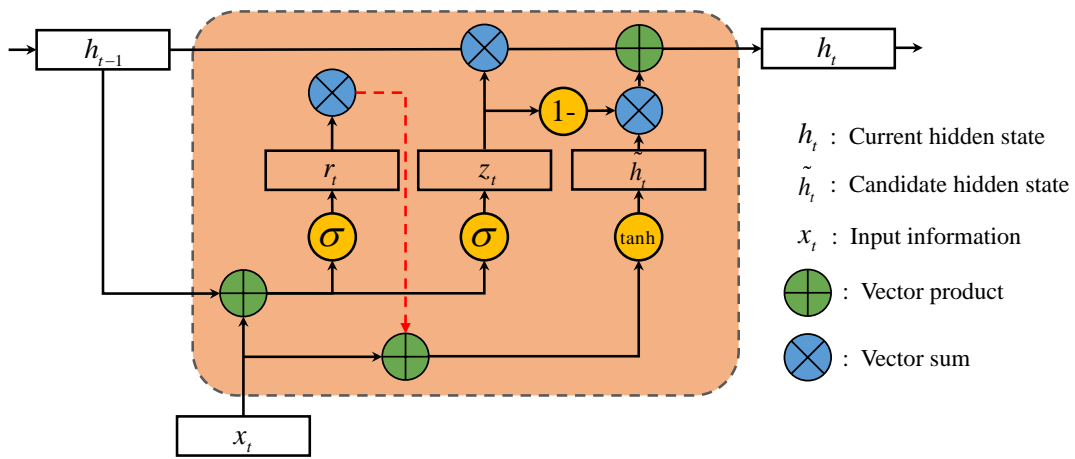
\plot{figure2}{width=1\textwidth}{The structure of GRU.}

\newpage
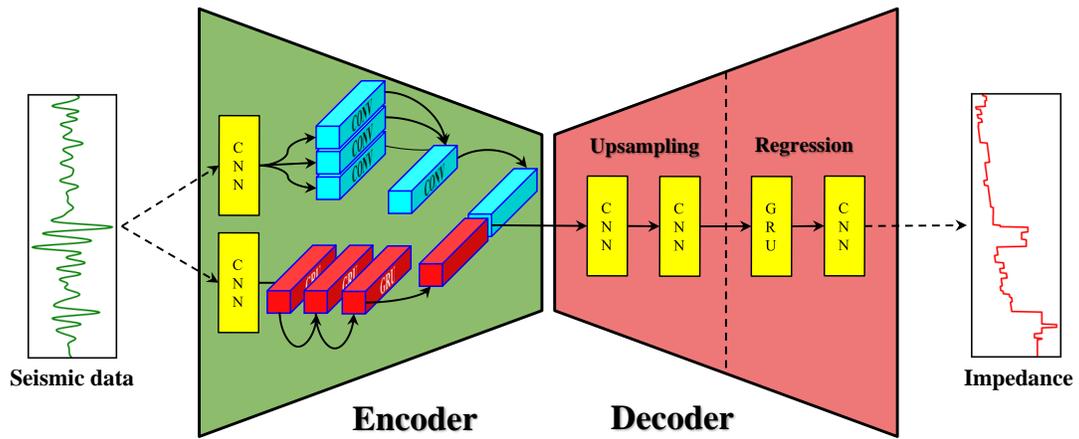
\plot{figure3}{width=1\textwidth}{Visualization of the Seismic Inversion module. The encoder consists of multiple GRU network blocks (blue) and multiple convolutional network blocks (red). The decoder consists of an upsampling layer and a regression layer, which is composed of a convolution layer and a GRU layer.}

\newpage
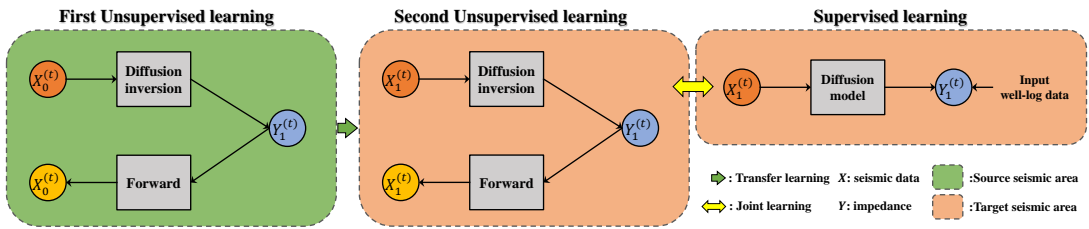
\plot{figure4}{width=\textwidth}{The Structure of unsupervised and semi-supervised joint learning.}

\newpage
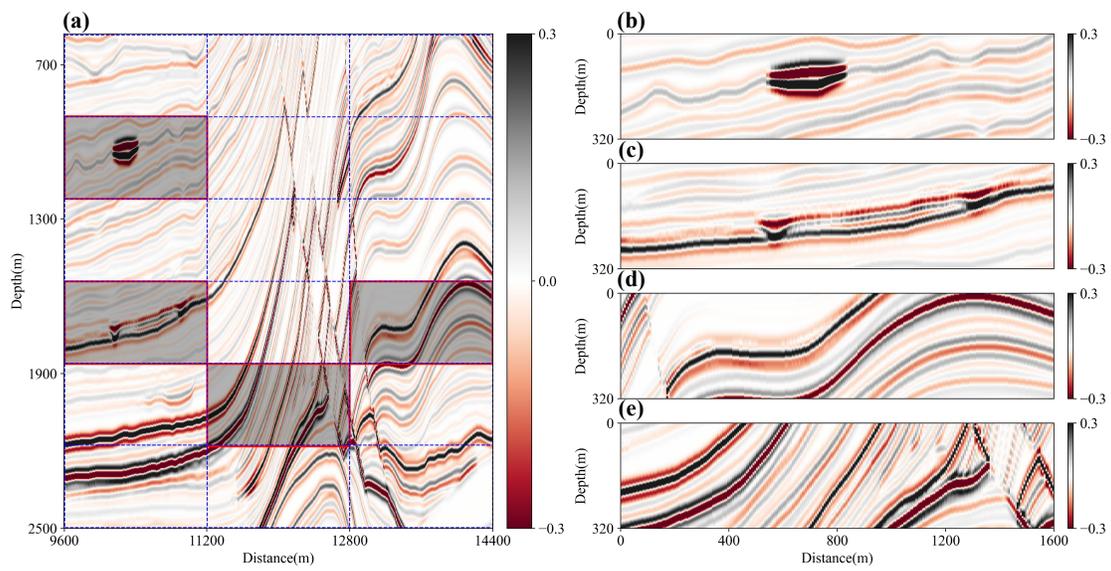
\plot{figure5}{width=\textwidth}{Segmentation processing of Marmousi2 data: (a) the whole seismic data, (b) seismic profile 1, (c) seismic profile 2, (d) seismic profile 3, and (e) seismic profile 4.}

\newpage
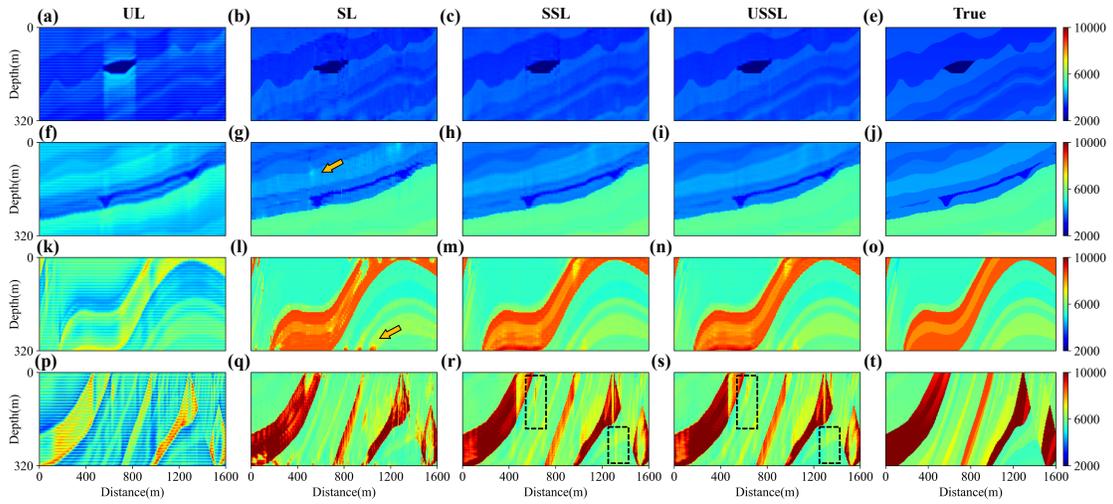
\plot{figure6}{width=\textwidth}{The impedance inversion results of different training methods in (a) - (e) seismic profile 1, (f) - (j) seismic profile 2, (k) - (o) seismic profile 3, and (p) - (t) seismic profile 4. The first column is UL, the second column is SL, the third column is SSL, and the last column is USSL.}

\newpage
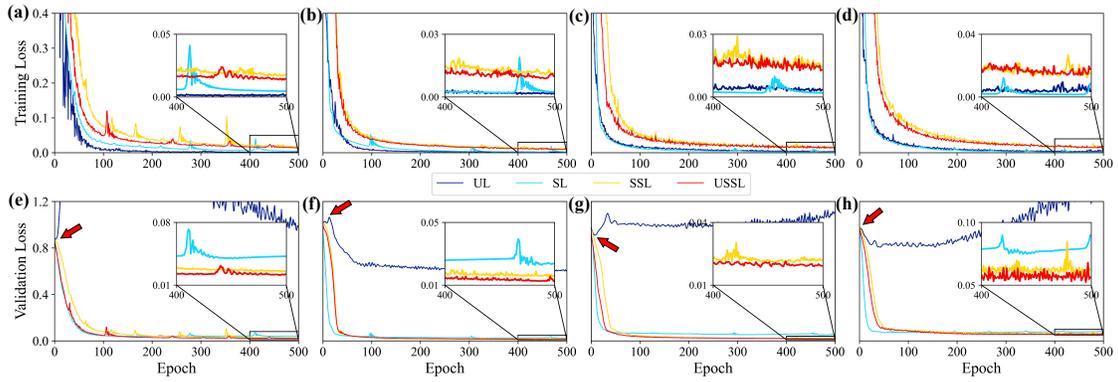
\plot{figure7}{width=\textwidth}{Loss function curves: (a)-(d) training data, (e)-(h) Loss function curves of validation data. The first column is the loss of seismic profile 1, the second column is the loss of seismic profile 2, the third column is the loss of seismic profile 3, and the last column is the loss of seismic profile 4. The blue line is UW, the sky blue line is SL, the orange line is SSL, and the red line is USSL.}

\newpage
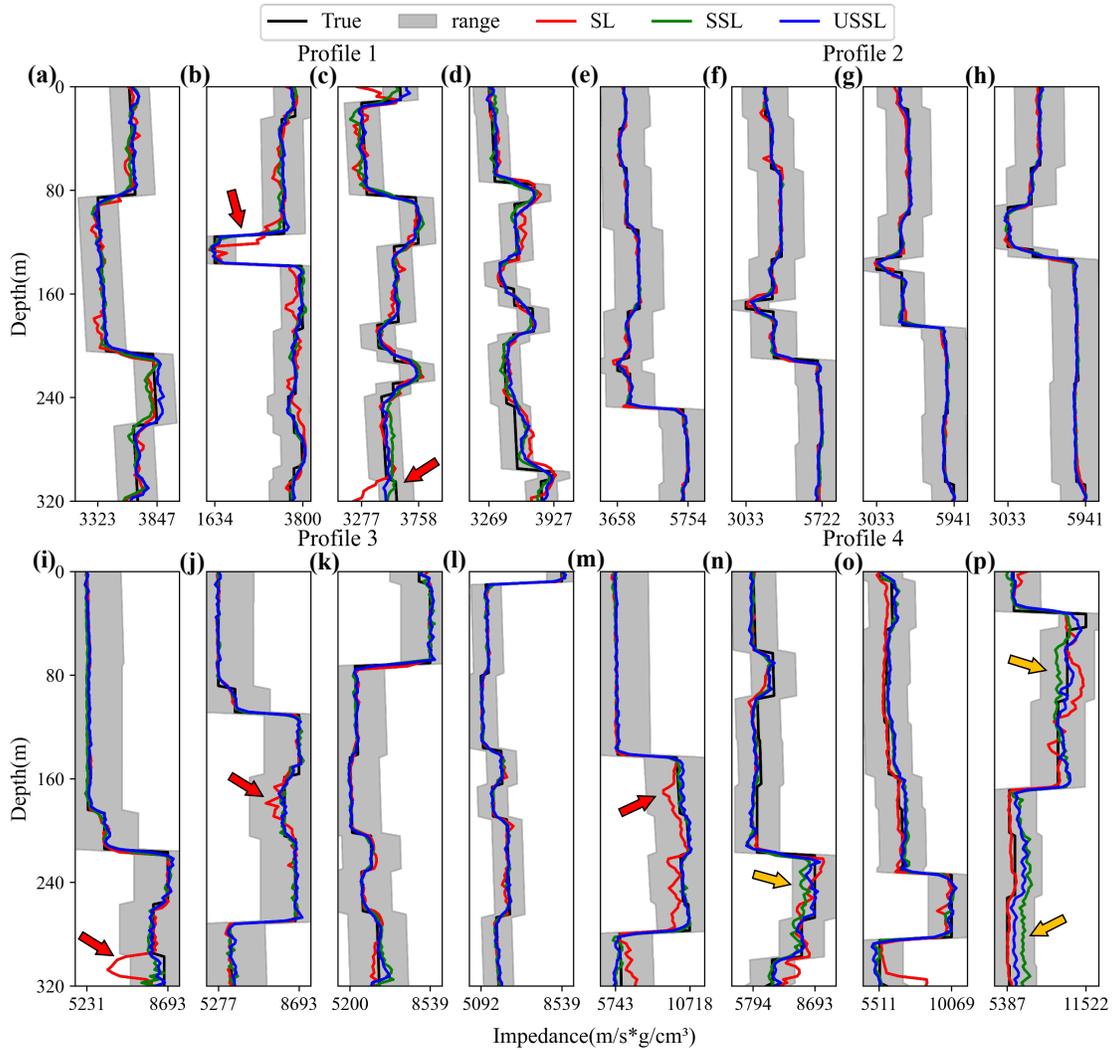
\plot{figure8}{width=\textwidth}{Single trace inversion results in (a) - (d) seismic profile 1, (e) - (h) seismic profile 2, (i) - (l) seismic profile 3 and (m) - (p) seismic profile 4. Each seismic profile contains inversion results of four single traces with distance =250m, distance =785m, distance =1060m, and distance =1310m. the black line is true data, the gray fill is the range of standard deviations of the true data, the orange line is the result of UL, the red line is the result of SL, the green line is the result of SSL, and the blue line is the result of USSL.}

\newpage
\plot{figure9}{width=\textwidth}{Inversion results of different methods in (a) seismic profile 1, (b) seismic profile 2, (c) seismic profile 3, and (d) seismic profile 4. The first line is seismic data with noise, the second line is unconditional SSL method inversion results, the third line is NSP-SSL method inversion results, the fourth line is RSP-SSL method inversion results, and the last line is DSIM-USSL method inversion results.
}

\newpage
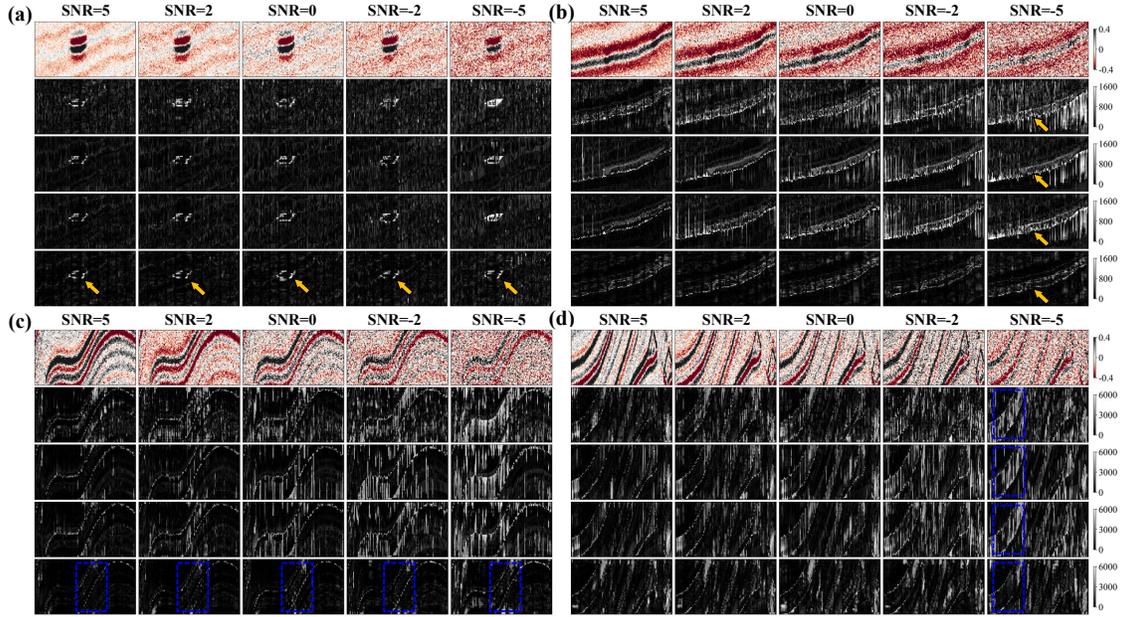
\plot{figure10}{width=\textwidth}{Absolute difference of inversion results of different methods in (a) seismic profile 1, (b) seismic profile 2, (c) seismic profile 3, and (d) seismic profile 4. The first line is seismic data with noise, the second line is the absolute difference of unconditional SSL method inversion results, the third line is the absolute difference of NSP-SSL method inversion results, the fourth line is the absolute difference of RSP-SSL method inversion results, and the last line is the absolute difference of DSIM-USSL method inversion results.
}

\newpage
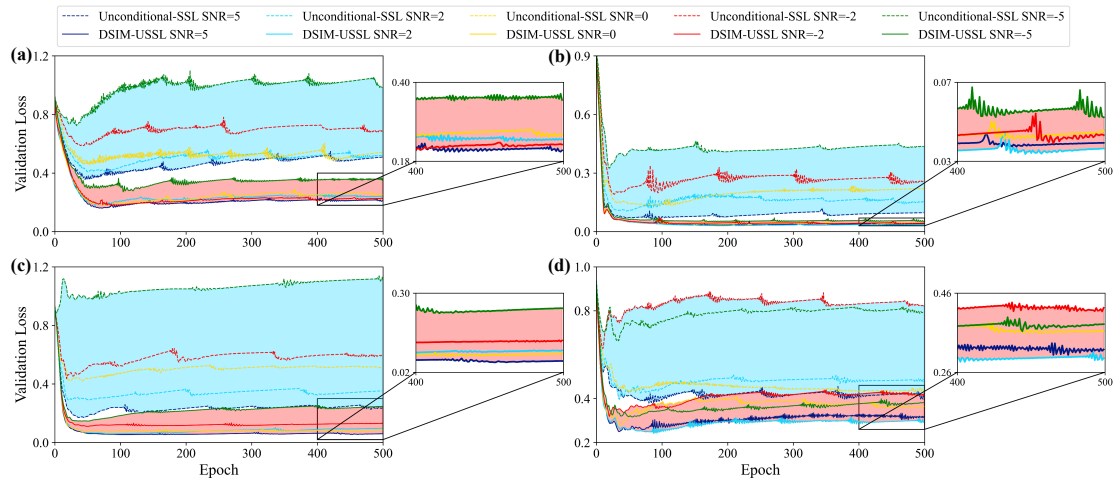
\plot{figure11}{width=\textwidth}{Comparison of validation loss function between unconditional SSL and DSIM-USSL in (a) seismic profile 1, (b) seismic profile 2, (c) seismic profile 3, and (d) seismic profile 4. The dashed line is the SSL method, and the solid line is the DSIM-USSL method. The blue fill is the unconditional SSL loss function range, and the red fill is the DSIM-USSL loss function range.}

\newpage
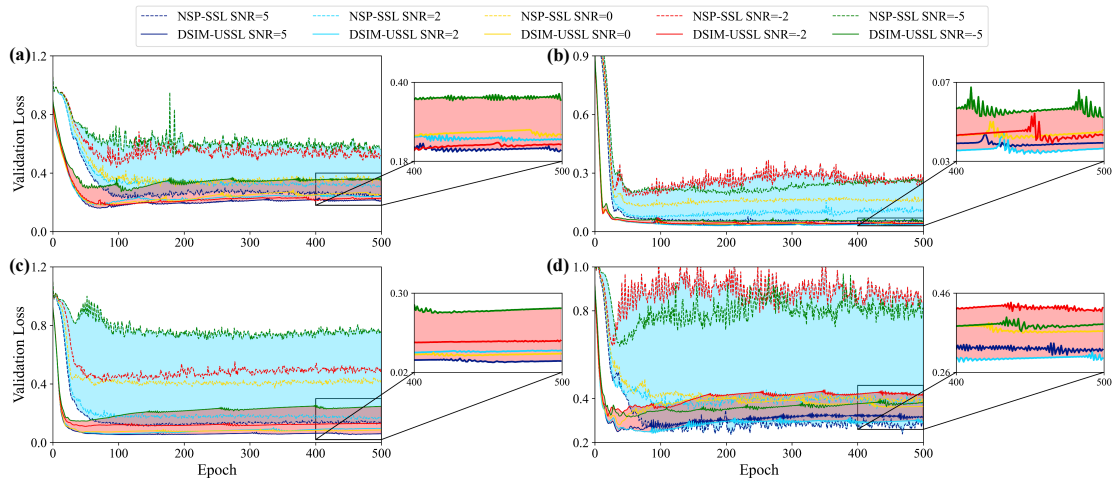
\plot{figure12}{width=\textwidth}{Comparison of validation loss function between NSP-SSL and DSIM-USSL in (a) seismic profile 1, (b) seismic profile 2, (c) seismic profile 3, and (d) seismic profile 4. The dashed line is the NSP-SSL method, and the solid line is the DSIM-USSL method. The blue fill is the NSP-SSL loss function range, and the red fill is the DSIM-USSL loss function range.}

\newpage
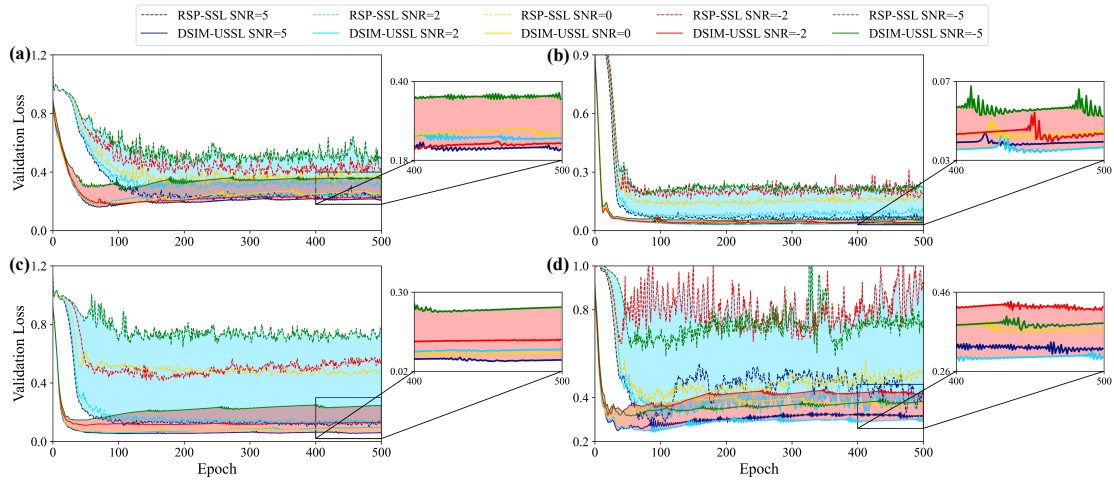
\plot{figure13}{width=\textwidth}{Comparison of validation loss function between RSP-SSL and DSIM-USSL in (a) seismic profile 1, (b) seismic profile 2, (c) seismic profile 3, and (d) seismic profile 4. The dashed line is the RSP-SSL method, and the solid line is the DSIM-USSL method. The blue fill is the RSP-SSL loss function range, and the red fill is the DSIM-USSL loss function range.}

\newpage
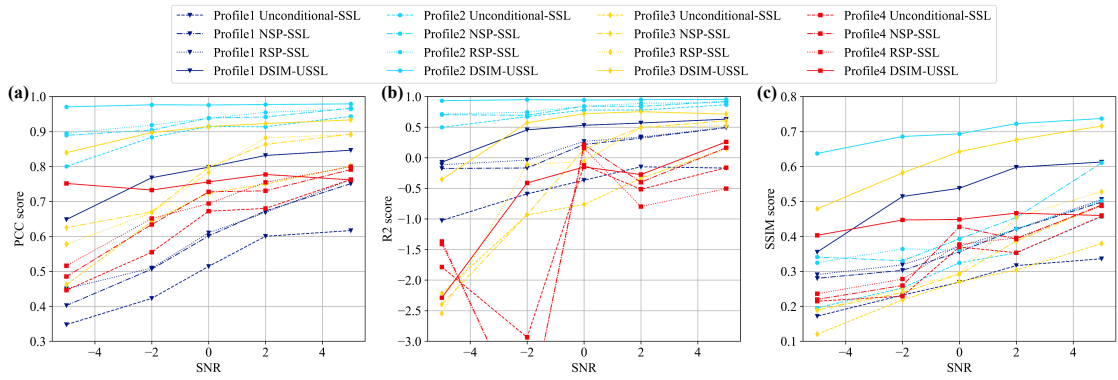
\plot{figure14}{width=\textwidth}{Curves of (a)PCC, (b)R2, and (c)SSIM evaluation indices. The dotted line is the unconditional SSL method, the dotted line is the NSP-SSL method, the dotted line is the RSP-SSL method, and the solid line is the DSIM-USSL method. The blue line represents seismic profile 1, sky blue represents seismic profile 2, orange represents seismic profile 3, and red represents seismic profile 4.}

\newpage
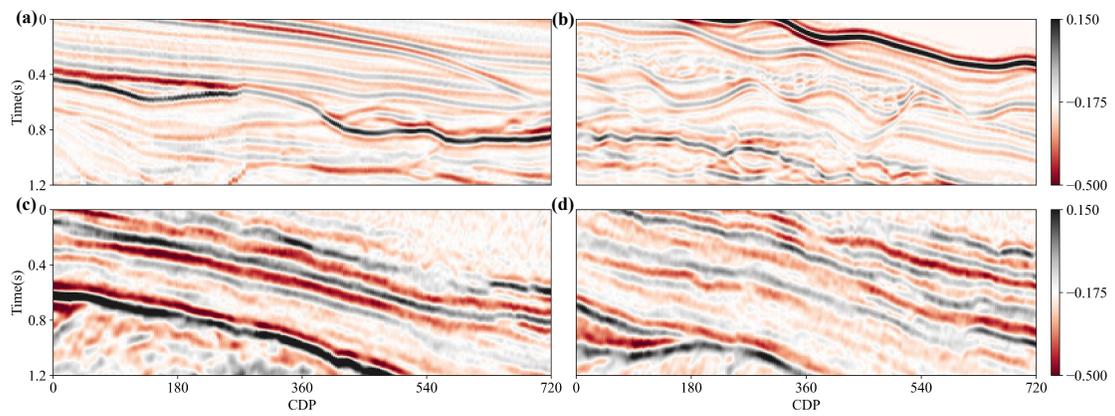
\plot{figure15}{width=\textwidth}{Seismic data without noise in (a) profile 1, (b) profile 2, (c) profile 3, and (d) profile 4 in Field H.}

\newpage
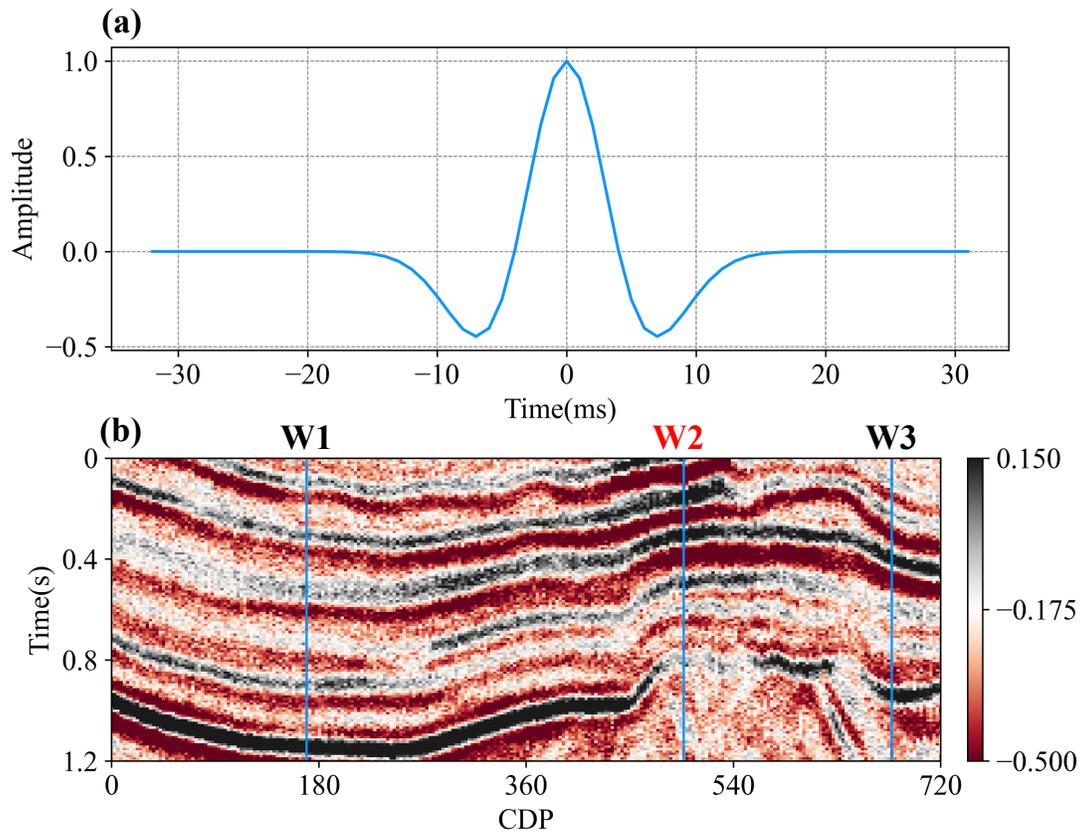
\plot{figure16}{width=\textwidth}{Properties of the target seismic area: (a) theoretical seismic wavelet and (b) post-stack seismic data. The solid blue lines in the figure below represent the three well-log locations with CDP values of 60,174 and 240, respectively.}

\newpage
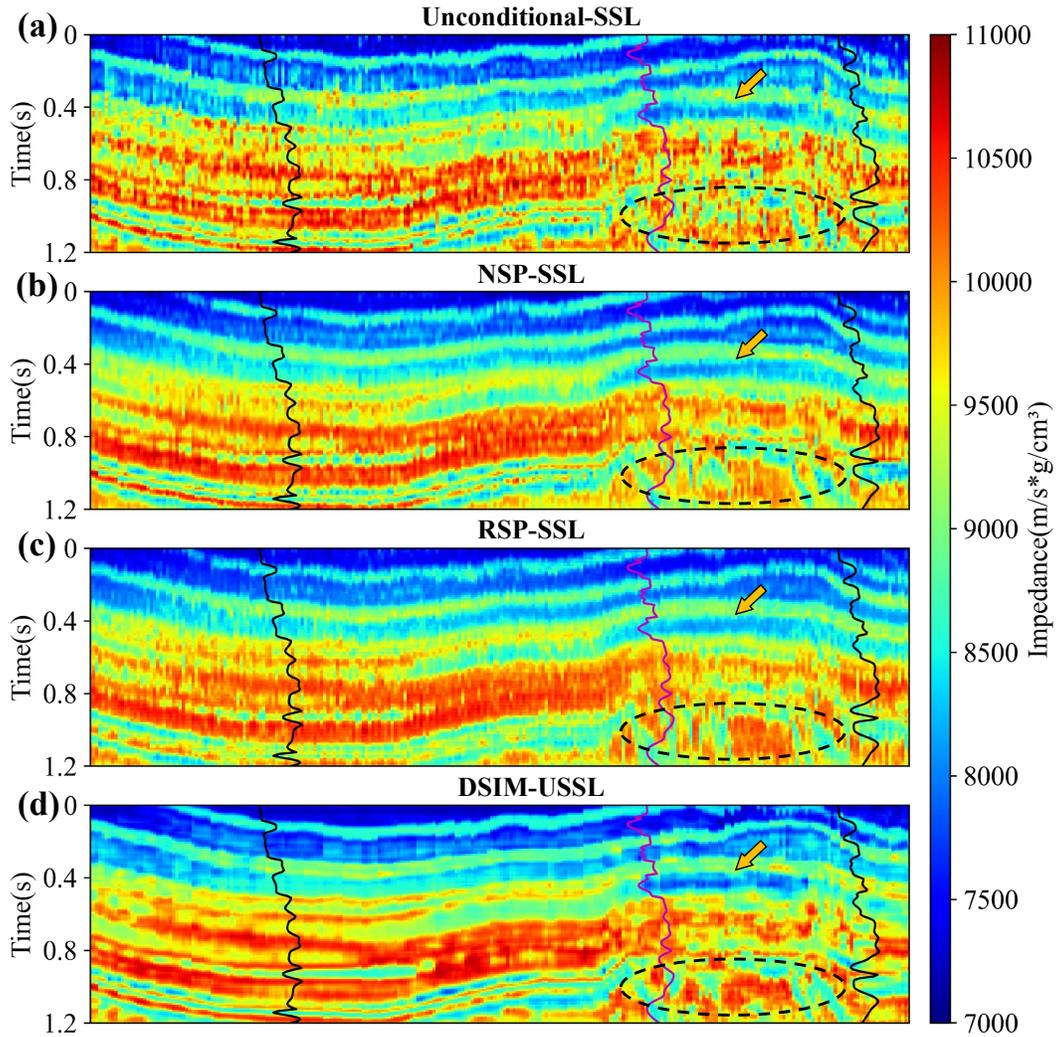
\plot{figure17}{width=\textwidth}{Impedance inversion results of (a) unconditional SSL method, (b) NSP-SSL method, (c) RSP-SSL method, and (d) DSIM-USSL method. The orange arrow indicates the gas reservoir. The oval wire frame indicates the sandstone region. The black curve is Wells W1 and W3 for training, and the pink curve is Well W2 for testing.}

\newpage
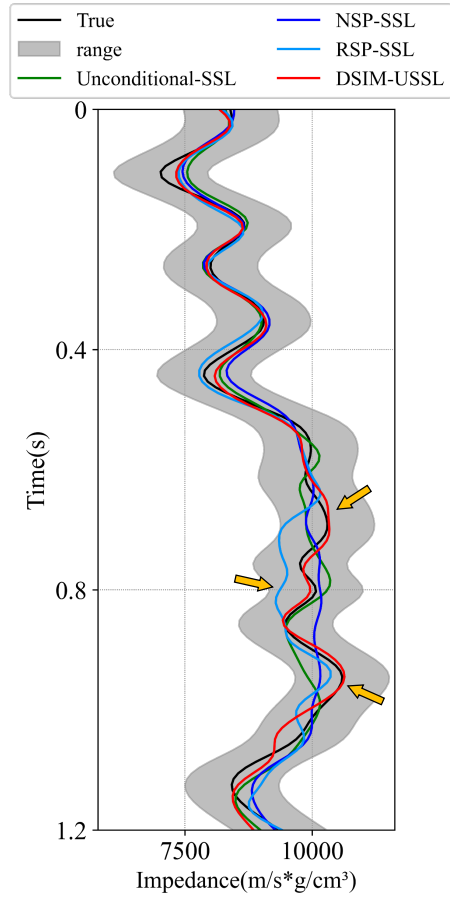
\plot{figure18}{width=0.5\textwidth}{Inversion results of wave impedance at the location of the well W2 on the field data. The black line indicates true data. The gray fill area indicates the range of standard deviations of the true data. The green line indicates the unconditional SSL method, the blue line indicates the NSP-SSL method, the sky blue line indicates the RSP-SSL method, and the red line indicates the DSIM-USSL method.}
\end{document}